\documentclass[fleqn,10pt]{wlscirep}
\usepackage[utf8]{inputenc}
\usepackage[T1]{fontenc}
\usepackage{tikz}
\usetikzlibrary{quantikz}

\title{Optimizing quantum noise-induced reservoir computing for nonlinear and chaotic time series prediction}

\author[1,*]{Daniel Fry}
\author[1]{Amol Deshmukh}
\author[2]{Samuel Yen-Chi Chen}
\author[1]{Vladimir Rastunkov}
\author[2]{\\Vanio Markov}
\affil[1]{IBM Quantum, Thomas J. Watson Research Center, Yorktown Heights, New York, USA}
\affil[2]{Wells Fargo, 150 East 42 Street, New York, NY 10017, USA}

\affil[*]{primary email: daniel.fry@ibm.com, secondary email: danielfry1988@gmail.com}

\begin{abstract}
Quantum reservoir computing is strongly emerging for sequential and time series data prediction in quantum machine learning. We make advancements to the quantum noise-induced reservoir, in which reservoir noise is used as a resource to generate expressive, nonlinear signals that are efficiently learned with a single linear output layer. We address the need for quantum reservoir tuning with a novel and generally applicable approach to quantum circuit parameterization, in which tunable noise models are programmed to the quantum reservoir circuit to be fully controlled for effective optimization. Our systematic approach also involves reductions in quantum reservoir circuits in the number of qubits and entanglement scheme complexity. We show that with only a single noise model and small memory capacities, excellent simulation results were obtained on nonlinear benchmarks that include the Mackey-Glass system for 100 steps ahead in the challenging chaotic regime.
\end{abstract}

\begin{document}

\flushbottom
\maketitle
%
%
\thispagestyle{empty}

\section*{Introduction}\label{sec:intro}

Quantum computing and machine learning have come together in recent years as a rapidly growing, cross-disciplinary field with huge transformative potential via quantum advantage. This new field of quantum machine learning (QML) aims to further revolutionize many of the areas that are currently seeing real transformation from new approaches in machine learning and artificial intelligence. An area that is critical to business and research is predicting or forecasting sequential time series data, which is paramount in finance, business, economics, climatology, meteorology, and ecology.

Reservoir computing (RC) is a paradigm for time series prediction that draws from some of the successful properties of RNNs, such as sequential memory, while greatly improving learning efficiency by fixing reservoir weights for all but a single trainable output layer \cite{jaeger2001echo,jaeger2004harnessing,maass2002real, lukovsevivcius2009reservoir, Mujal2021Opportunities, tanaka2019recent}. While RC is well-suited to dynamical system modeling, it is proven to be a universal approximator for sequential functions \cite{dambre2019information}. The quantum-enhanced version of RC (QRC) leverages a quantum reservoir, a natural quantum many-body system or a programmable quantum computer circuit. QRC provides a path to quantum advantage by using a quantum reservoir with an exponentially larger computational space and greater complexity for time series prediction.

Many QRC frameworks have been developed, under which QRC algorithms can be classified at a higher level. For example, QRC reservoir nodes have been realized with quantum basis states \cite{markovic2020quantum, govia2021quantum} in contrast to qubits or qudits. One such novel framework leverages hybrid quantum-classical RNNs, such as quantum long short-term memory (QLSTM), as a reservoir \cite{chen2022reservoir}. QRC reservoirs can be based directly on known quantum system models or on hardware-efficient quantum feature map designs, where this work falls under the latter. In this work we focus on noisy or dissipative quantum reservoirs, pioneered in works \cite{suzuki2022natural, kubota2022quantum, domingo2023taking}. Furthermore, we demonstrate a successful, novel approach to quantum reservoir optimization, where other schemes have been explored in works \cite{kutvonen2020optimizing, dasgupta2020designing, domingo2022optimal}.

QRC has been applied to many prediction tasks including nonlinear time series prediction \cite{chen2022reservoir, suzuki2022natural}, time series classification \cite{suzuki2022natural}, image recognition \cite{burgess2022quantum}, and stock market value \cite{kutvonen2020optimizing} and volatility prediction \cite{dasgupta2020designing}. In quantum information science, QRC has been used for entanglement recognition, non-linear function estimation and quantum state tomography \cite{ghosh2019quantum, ghosh2020reconstructing, angelatos2021reservoir}.

QRC may have begun to demonstrate superior computational capacity to classical RC. In one example, numerical studies have shown that quantum reservoirs consisting of 5–7 qubits possess computational capacities comparable to conventional recurrent neural networks of 100-500 nodes \cite{fujii2017harnessing}. In this work we demonstrate excellent prediction capacity of few-qubit reservoirs. 

The most significant computational capacity that is a main goal of QML is quantum advantage, which is a measurable performance improvement over classical computation on a well-defined objective task (e.g. a business time series prediction task) using quantum computation \cite{bravyi2022future}. Quantum advantage with QRC likely exists if the quantum reservoir requires a complex, many-qubit entangled architecture that is intractable to classical computation. This view is the same as that expressed in \cite{havlivcek2019supervised} for quantum-enhanced feature spaces, a closely related approach to QRC, where the quantum reservoir acts as a sequential feature map.
\\
\\
In this work we build on the quantum noise-induced reservoir (QNIR) framework \cite{suzuki2022natural, kubota2022quantum}, with a novel approach to parameterized quantum circuits for the reservoir and a systematic reduction of circuit complexity. The term \textit{reduction} is used for minimizing quantum circuit resources to clearly differentiate it from the reservoir optimization achieved with parameterized noise channels. QNIR is a type of QRC that relies on quantum hardware noise or, as in the focus of this work, artificial noise models in quantum software, as a resource to generate rich, dissipative quantum reservoir dynamics. In the current, transitional NISQ phase of quantum computing, QNIR can use inherent hardware noise. However, in future strongly error-mitigated and fault-tolerant quantum computers, QNIR noise channels can be coded instructions in a quantum program along with quantum gates. This approach abstracts this QNIR algorithm from the underlying physical device. In a novel approach we implement parameterized artificial noise models programmed to a quantum computer for improved time series prediction performance. With this, we address the important need of reservoir tuning, in QNIR and QRC in general, for classes of prediction tasks.

Powerful optimization approaches for reservoir noise are offered by dual annealing and evolutionary optimization (EO). EO is capable of optimizing quantum systems at various levels, such as quantum circuit parameters, successfully realized in this work and in previous works \cite{anand2020natural,franken2022quantum,lu2020markovian} and quantum circuit architecture. Here we use a previously successful EO algorithm \cite{chen2022variational} in which model parameters were evolved for quantum reinforcement learning agents in a hybrid quantum-classical neural network approach.
\\
\\
\section*{Quantum noise-induced reservoir computing} \label{sec:qnir_computing}

\subsection*{Theoretical framework}\label{subsection:QNIR_Theory}

We develop QNIR theory starting from general RC theory. RC is a computational paradigm and class of machine learning algorithms that derives from RNNs. RC involves mapping input signals, or time series sequences, into higher dimensional feature spaces provided by the dynamics of a non-linear system with fixed coupling constants, called a reservoir. Having a smaller number of trainable weights confined to a single output layer is a core benefit of RC because it makes training fast and efficient compared to RNNs. RC has a number of properties that should be met \cite{dasgupta2022characterizing, martinez2023quantum} including adequate reservoir dimensionality, nonlinearity, fading memory/echo state property (ESP) and response separability.

For the univariate case, a reservoir, $f$, is a recurrent function of an input sequence, $u_t$, and prior reservoir states, $\bar{x}_{t-1}$, as
\begin{equation}
\bar{x}_t = f(\bar{x}_{t-1},u_t).
\label{eqn:QRC map}
\end{equation}

As output sequences, $\bar{x}_t$, training sequences are selected between time-steps $t=t_i$ and $t=t_f$, and form a training design matrix, $\mathbf{X}_{tr}$. The initial sequence, $t<t_i$, is a washout interval required required for fading memory/ESP. A multiple linear regression model with an initial form: 
\begin{equation}
\mathbf{y} = W^T \mathbf{X}_{tr},
\label{eqn:linear regression training}
\end{equation}
is trained based on least squares, where $\mathbf{y}$ is the target vector and $W$ is an initial weight vector. The trained model has the form:
\begin{equation}
\hat{\mathbf{y}} = W^T_{opt}\mathbf{X},
\label{eqn:linear regression}
\end{equation}
with an optimized weight vector, $W^T_{opt}$, to give a predicted sequence, $\hat{\mathbf{y}}$, from new sequences, $\mathbf{X}$.

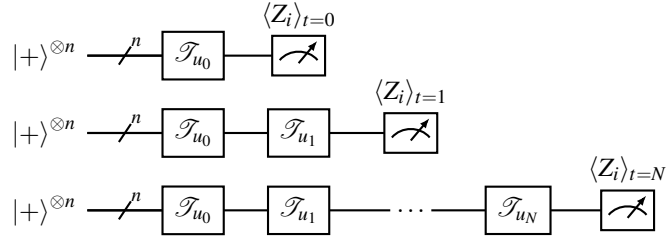
\begin{figure}[h!]
\centering
\begin{quantikz}[row sep=0.0cm]
\lstick{$\ket{+}^{\otimes n}$} & [5mm] \gate{\mathcal{T}_{u_0}}\qwbundle
{n} & \meter{$\langle Z_{i} \rangle_{t=0}$}
\\
\lstick{$\ket{+}^{\otimes n}$} & [5mm] \gate{\mathcal{T}_{u_0}}\qwbundle
{n} & \gate{\mathcal{T}_{u_1}} & \meter{$\langle Z_{i} \rangle_{t=1}$}
\\
\lstick{$\ket{+}^{\otimes n}$} & [5mm] \gate{\mathcal{T}_{u_0}}\qwbundle
{n} & \gate{\mathcal{T}_{u_1}} &  \ \ldots\ \qw &
\gate{\mathcal{T}_{u_N}} & \meter{$\langle Z_{i} \rangle_{t=N}$}
\end{quantikz}
\caption{Circuit channel diagrams of the QNIR computer in the unrolled view, composed using \cite{kay2018tutorial}. The initial state of the quantum reservoir is $\ket{+}^{\otimes n}$ and the quantum channels labeled $\mathcal{T}_{u_i}$ evolve the density operator as in Eq. \ref{eqn:QRC evolution}, where $N$ quantum circuits are required for $N$ time steps. A number of output sequences, $n$, are concatenated from sequential, single-qubit expectation value measurements $\langle Z_{i} \rangle$ on $n$ qubits.}
\label{fig:QNIR_figure}
\end{figure}

For QNIR with artificial noise channels, the RC framework that has been developed is now instantiated in the following way. The density operator evolves in time steps as 
\begin{equation}
\rho_t = \mathcal{T}_{u_t}(\rho_{t-1}),
\label{eqn:QRC evolution}
\end{equation}
where the reservoir map $\mathcal{T}_{u_t}$ is composed of a sequence unitary quantum gates, $U_i$, and associated artificial noise channels, $\mathcal{E}_i$, that are completely positive and trace preserving (CPTP). The reservoir map can be represented as a composition of quantum channels
\begin{equation}
\mathcal{T}_{u_t}(\rho_{t-1}) = \mathcal{E}_{U_K} \circ \ldots \circ \mathcal{E}_{U_2} \circ \mathcal{E}_{U_1} (\rho_{t-1}),
\end{equation}
where the notation $\mathcal{E}_{U_i} = \mathcal{E}_i( U_i \rho U_i^{\dagger} )$ is used for clarity and to emphasize that each quantum gate is acted on by a noisy channel and $K$ is the number of noise channels in the time step. We will refer to $\mathcal{T}_{u_t}$ as a noisy quantum circuit. QNIR requires an initial washout phase, $t<t_i$, where the reservoir forgets its initial state before a steady state is reached.

The unitary, noiseless part of the quantum circuit is composed of an initial layer of $RX$ gates followed by an entanglement scheme of ${RZ\!Z}_{i,j}$ gates, which are 2-qubit entangling gates
\begin{equation} (C\!X_{i,j}RZ_j(\theta)C\!X_{i,j})RX^{\otimes n}(\theta) =  {RZ\!Z}_{i,j}(\theta)RX^{\otimes n}(\theta),
\label{eqn:QNIR encoding circuit general}
\end{equation}
where all $RX(\theta)$ and $RZ(\theta)$ rotation gates encode the time series data with a scaling map, $\theta=\phi(u)$. The purpose and structure of the unitary encoding gates is detailed in subsection: Reservoir circuit designs.

\begin{figure}[h!]
\centering
\includegraphics[width=10cm]{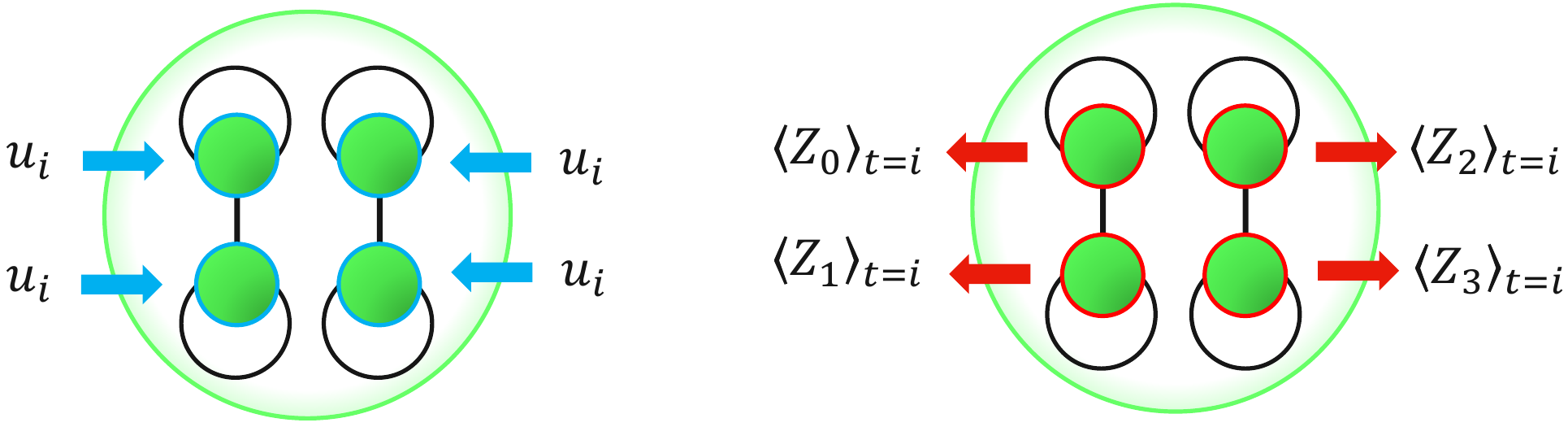}
\caption{This drawing represents many repeats  of data encoding of a single value, $u_i$, to all reservoir qubits (left) and measurements of single-qubit $Z$ expectation values (right). This two-part process occurs at each time step $i$ to build feature signals by concatenation. Noisy quantum circuits are shown for each time step in Fig. \ref{fig:QNIR_figure}. This drawing shows an example of a four-qubit reservoir with fixed, pair-separable dynamics.}
\label{fig:RC process}
\end{figure}

\begin{figure}[h!]
\centering
\includegraphics[scale=0.45]{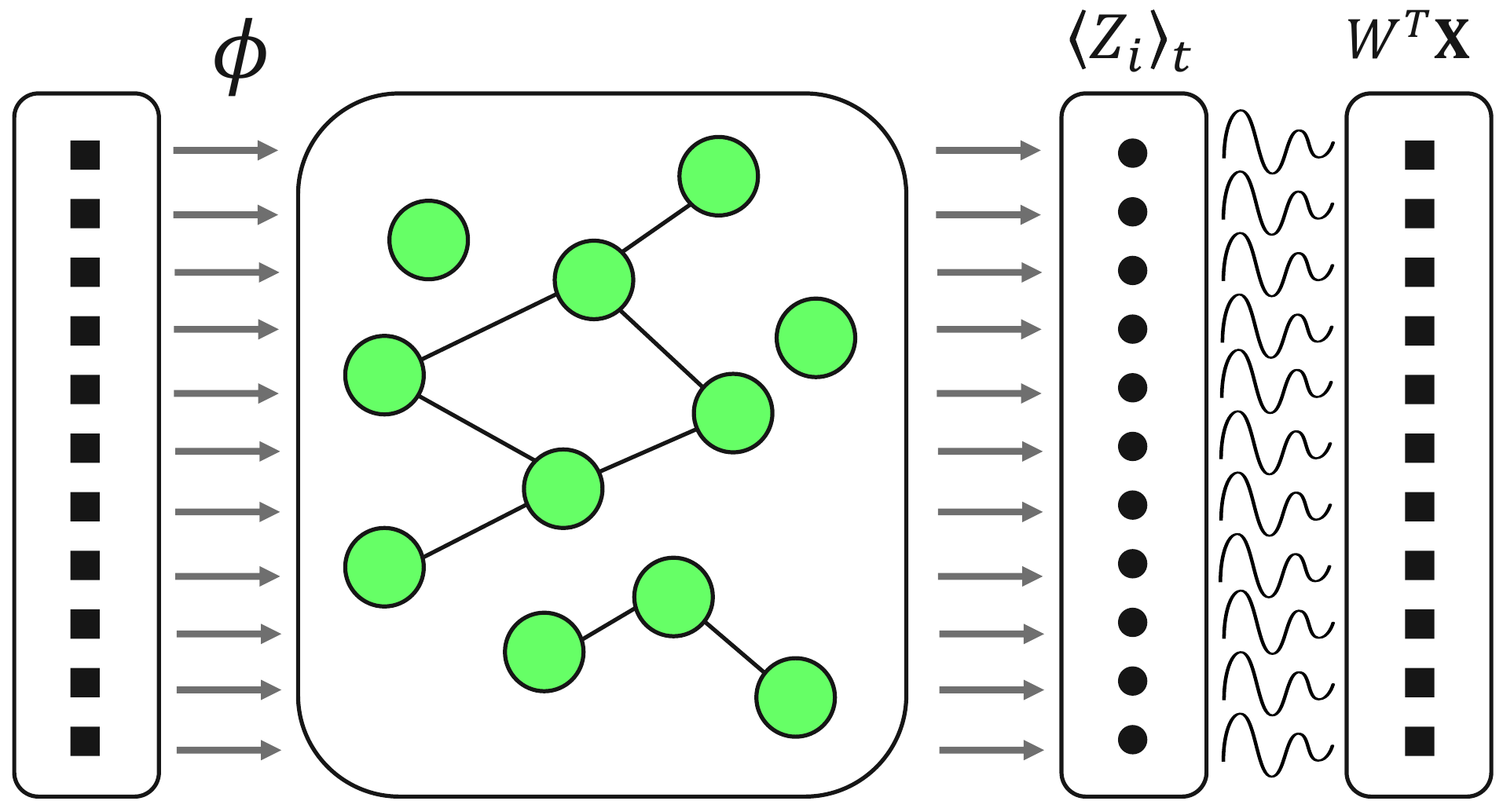}
\caption{In this graphic the first layer contains an array of duplicates of a single time series value. Each value in the input array is encoded to all qubits of the reservoir as in Eq. \ref{eqn:QNIR encoding circuit general}. The second layer is a quantum reservoir with arbitrary entanglement scheme, represented by connecting lines between qubit nodes. The $Z$ observable expectation value, $\langle Z_{i}\rangle$, is measured for all qubits. These measurements are repeated and concatenated to build output signals, $q_i$. In the final layer, these signals are used in multiple linear regression for time series prediction, as in Eq. \ref{eqn:linear regression}.}
\label{fig:QNIR scheme}
\end{figure}

Single-qubit expectation values, $\langle Z_{i} \rangle = Tr(Z_i \rho)$, are measured for all $n$ qubits at each time-step,
\begin{equation}
h_t = [\langle Z_{1} \rangle,\langle Z_{2} \rangle,\ldots,\langle Z_{n} \rangle]^T,
\end{equation}
as shown in a circuit diagram in Fig. \ref{fig:QNIR_figure}. Fig. \ref{fig:RC process} depicts that time series values are encoded to all reservoir qubits and $\langle Z_{i} \rangle$ are measured of all qubits, which are concatenated for each time step to give $n$ reservoir feature sequences $q_i = \{\langle Z_{i} \rangle\}_{t=0}^N$, where $N$ is the number of time steps. In turn, $q_i$ form a design matrix $\mathbf{X}$ and the QNIR model is trained as in Eq. \ref{eqn:linear regression}. A schematic of the full QNIR computer is shown in Fig. \ref{fig:QNIR scheme}.

It is important in RC and by extension QRC that the reservoir system can capture the temporal dynamics of the target system. To ensure this we implement a reservoir optimization scheme for QNIR. The artificial noise channels, $\mathcal{E}_i$, of the quantum reservoir circuit are iteratively updated by an optimization routine with an MSE cost function based on the time series prediction performance. This serves to optimize the quantum reservoir for time series prediction. Details of the optimization approach are in subsection: Reservoir noise parameterization.

\subsection*{Reservoir circuit designs}
\label{subsection:Reservoir circuit designs}
This section is concerned with the  architecture and purpose of the unitary gates of the quantum circuit, the high-level structure of the noisy quantum circuits and entanglement scheme. The details of the noise scheme are covered in subsection: Reservoir noise parameterization. 

The initial state of the quantum reservoir, $|+\rangle^{\otimes n}$, is prepared by an initial Hadamard gate layer. Continuing with Eq. \ref{eqn:QNIR encoding circuit general}, an $n$-qubit QNIR circuit has a fixed sequence of quantum gates
\begin{equation}
\begin{split}
    U_{b}(u) &= (C\!X_{i,j}RZ_j(\theta)C\!X_{i,j})RX^{\otimes n}(\theta) \\
    &= {RZ\!Z}_{i,j}(\theta)RX^{\otimes n}(\theta)
\end{split}
\label{eqn:QNIR encoding circuit general2}
\end{equation}
where $i,j$ are indices for two qubits that denote the placement of multiple 2-qubit $RZ\!Z$ entangling gates. The decomposed form of the circuit with $C\!X$ and $RZ$ gates \cite{havlivcek2019supervised} is implemented with noise channels (see subsection: Reservoir noise parameterization). A time series data value, $u$, is encoded to all $RX(\theta)$ and $RZ\!Z(\theta)$ gates by angle $\theta = \phi(u)$, where $\phi$ is a scaling map.

To implement the recurrent architecture of QNIR, a set of $N$ quantum circuits are executed for a time series $\{u_t\}^N_{t=0}$. The first circuit encodes $\{u_0\}$, the second circuit encodes $\{u_0,u_1\}$, and the $N$th circuit encodes $\{u_t\}^N_{t=0}$ as
\begin{equation}
\text{U}_{t=N} = U_{b}(u_N) \ldots U_{b}(u_1)U_{b}(u_0).
\label{eqn:unitary_sequence}
\end{equation}

All unitaries $\text{U}_t$ for arbitrary $t$ constrain the $i$ expectation values to a zero bitstring
\begin{equation}
\langle Z_{i} \rangle_{t} = 
\bra{\Phi_0}\text{U}^{\dagger}_t Z_i \text{U}_t \ket{\Phi_0} = 000...,
\end{equation}
where $\ket{\Phi_0} = \ket{+}^{\otimes n}$ is the initial reservoir state and $Z_i$ represents $n$ single-qubit $Z$ measurement operators. It is the action noise that ensures the qubit signals are non-zero, feature sequences, $q_i$. Now considering the full QNIR circuits with artificial noise, the noisy quantum circuit for the final iteration, encoding $\{u_t\}^N_{t=0}$, is the quantum channel
\begin{equation}
\boldsymbol{\mathcal{{T}}}_{N} = 
\mathcal{T}_{u_N} \circ \ldots \circ \mathcal{T}_{u_2} \circ \mathcal{T}_{u_1}.
\end{equation}
The noisy quantum circuit with artificial noise scheme will be detailed in the next subsection: Reservoir noise parameterization. This scheme may further reduce resources by circuit truncation based on a memory criterion \cite{chen2020temporal, martinez2023quantum, vcindrak2023solving, mujal2023time}.

For $RZ\!Z_{i,j}$ gates, the degree of entanglement between qubits $i$ and $j$ is a function of $u_t$. It is important that the range of magnitudes of the data values is constrained and we observe that values much larger than $2\pi$ cause undesirable effects. We consider benchmarks that do not require re-scaling.

Drawing from the close connection with quantum feature maps, \cite{havlivcek2019supervised,schuld2019quantum, lloyd2020quantum, schuld2021machine} entanglement schemes are defined by the number and placement, i.e. the architecture, of $RZ\!Z$ gates in Eq. \ref{eqn:QNIR encoding circuit general}. Common entanglement schemes that could be trialed are full, linear, pair-wise, and what we call \emph{pair-separable} used in \cite{suzuki2022natural}. The pair-separable (PS) and linear entanglement (LE) schemes explored in this work have $RZ\!Z$ gates indexed as $i,j \in \{(0,1),(2,3),(4,5),...,(N-1,N)\}$ and respectively $i,j \in \{(0,1),(1,2),(2,3),...,(N-1,N)\}$. To clarify, for an LE scheme, every additional $RZ\!Z$ gate is in a new circuit layer, increasing the circuit depth each time. The LE scheme creates whole circuit entangled states \cite{havlivcek2019supervised}. The state vector for a PS entanglement scheme evolves in a product state of qubit pairs, $\ket{\psi} = \bigotimes_{i=1}^{n/2} \ket{\phi}_i$, where $\ket{\phi}_i$ are two-qubit entangled states. The state, $\ket{\psi}$, can be efficiently classically simulated and can be parallelized in classical simulation or on quantum computers \cite{nakajima2019boosting, tran2020higher}.

\subsection*{Reservoir noise parameterization}
\begin{figure}[h!]
\centering
    \begin{quantikz}[column sep=1.5mm]
    & \gate{RX(\theta)} & \gate{\mathcal{E}(p_0)}  & \ctrl{1} & \gate{\mathcal{E}(p_1)} & \qw & \qw & \ctrl{1} & \gate{\mathcal{E}(p_2)} & \qw\\
    & \gate{RX(\theta)} & \gate{\mathcal{E}(p_3)}  & \targ{} & \gate{\mathcal{E}(p_4)} &\gate{RZ(\theta)} & \gate{\mathcal{E}(p_5)} & \targ{} & \gate{\mathcal{E}(p_6)} & \qw
    \end{quantikz}
    \caption{A 2-qubit quantum circuit channel diagram of an reservoir noise parameterization. Each unitary gate has an associated noise channel represented by $\mathcal{E}(p_i)$. This represents the novel quantum circuit parameterization approach proposed in this work.}
\label{fig:Noisy quantum circuit}
\end{figure}
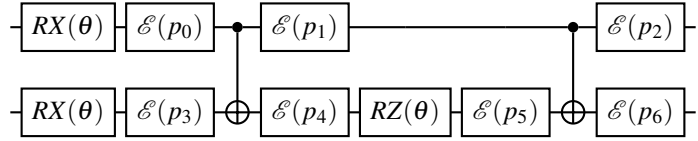

QNIR uses noise as a necessary resource to generate non-trivial feature sequences. We use artificial noise that can be programmed to a quantum computer. Within this scheme, many such artificial noise models can be implemented to produce different effects. To implement a noise scheme, we associate parameterized, single-qubit noise channels with each unitary gate in the quantum circuit, Eq. \ref{eqn:QNIR encoding circuit general}, as shown in Fig. \ref{fig:Noisy quantum circuit}. Note that this differs from Kubota et al.\cite{kubota2022quantum}, where noise channels were situated at the end of every time step. In the following, we assume each noise channel depends on a single noise parameter.

\begin{figure*}
\centering
\includegraphics[scale=0.55]{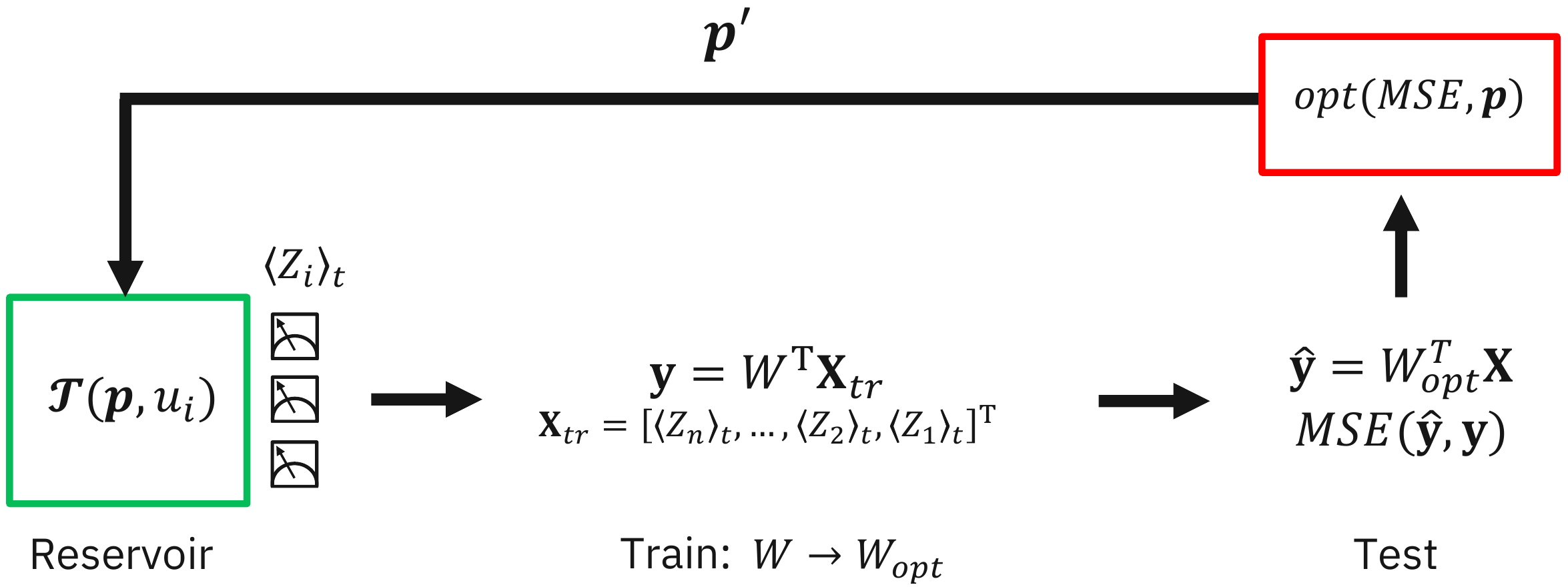}
\caption{This graphic shows the QNIR noise optimization scheme. The quantum model is trained and tested iteratively in a classical optimization loop, where dual annealing or evolutionary optimization are used. The quantum reservoir circuits have a number of gate-associated noise channels, each of which has a single error probability parameter that is iteratively updated.}
\label{fig:QNIR optimization graphic}
\end{figure*}

Noise channels are associated with all quantum gates in the reservoir circuit in Fig. \ref{fig:Noisy quantum circuit}. Each noise channel $\mathcal{E}(p)$ is a function of a probability for the noise effect to occur. We use probabilities, $p_i$, to parameterize the reservoir for optimization. The number of probability parameters scales linearly with the number of qubits. For pair-separable entanglement reservoir, the number of parameters is $n_{p_i} = \frac{7}{2} n$, where $n=2,4,6,...$, and for linear entangled reservoir $n_{p_i} = 6n-5$, where $n=2,3,4,...$.

QNIR resource-noise optimization is performed through iterative training (Eq. \ref{eqn:linear regression training}) and testing (Eq. \ref{eqn:linear regression}) of QNIR, giving optimized noise probability parameters, $p_i \in \mathbf{p}$ (see Fig. \ref{fig:QNIR optimization graphic}). The parameters in the initial parameter vector, $\mathbf{p}$, are probabilities randomly selected from a uniform distribution, $p_i \sim U(0,1), \forall i$. 

Two optimization approaches were trialed in this work, evolutionary optimization \cite{chen2022variational} and dual annealing \cite{xiang1997generalized}, where the latter is available in the SciPy optimization package \cite{2020SciPy-NMeth}. The mean squared error (MSE) was used as a suitable cost function to measure prediction performance, which is minimized as
\begin{equation}
\min_{\mathbf{p}}\; \{ \text{MSE}(\hat{\mathbf{y}}(\mathbf{p}),\mathbf{y}) : p_i \in [0,1], \forall i \},
\end{equation}
where
$\hat{\mathbf{y}} = W^T_{opt} \mathbf{X}(\mathbf{p})$ is the QNIR test set prediction and $\mathbf{X}(\mathbf{p})$ are the reservoir signals matrix dependent on noise probabilities $\mathbf{p}$.

In this work, we use only reset noise channels that can be simply implemented with a classical ancilla system (see next subsection: Reset noise).

\subsection*{Reset noise}
We propose a simple hybrid quantum-classical algorithm for a reset noise channel that consists of probabilistically triggering a reset instruction using a classical ancillary system. A deterministic reset instruction is an important element of a quantum instruction set, for the need to reset qubit states. A quantum instruction set is an abstract quantum computer model\cite{cross2021openqasm, smith2016practical}. In this work we consider a reset to $|0\rangle$ noise channel given by $\mathcal{E}_{PR}(\rho) = p|0\rangle\langle0| + (1-p)\rho$, where $p$ is the reset probability \cite{gutierrez2013approximation}. $\mathcal{E}_{PR}(\rho)$ is trace-preserving, $Tr(\mathcal{E}_{PR}(\rho))=1$.

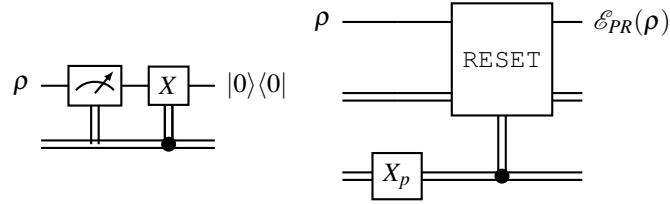
\begin{figure}[h!]
\centering
\begin{quantikz}[column sep=3.5mm]
\lstick{$\rho$} & \meter{}\vcw{1} & \gate{X} \vcw{1} & \rstick{$|0\rangle\langle0|$} \qw\\
& \cw & \cwbend{-1} & \cw
\end{quantikz}
\hspace{1pt}
\begin{quantikz}[column sep=4mm]
\lstick{$\rho$} & \qw & \gate[2, cwires={2}]{\texttt{RESET}} & \rstick{$\mathcal{E}_{PR}(\rho)$} \qw \\
& \cw & & \cw \\
& \gate[cwires={1}]{X_p} & \cwbend{-1} & \cw
\end{quantikz}
\caption{A deterministic \texttt{RESET} instruction (left) is executed with this dynamic circuit. This can be used as a basis for a reset noise channel, $ \mathcal{E}_{PR} $. A single line represents a qubit and a double-line represents a classical bit. A model classical ancillary system (right) would be executed on a classical computer. The classical NOT gate, $X_p$, is executed with probability $p$, which in turn triggers a classical controlled \texttt{RESET} instruction with probability $p$.}
\label{fig:Reset noise}
\end{figure}

Using dynamic circuits, quantum computers can implement a reset instruction with a mid-circuit measurement followed by a classically controlled quantum $X$ gate that depends on the measurement outcome\cite{corcoles2021exploiting} (see Fig. \ref{fig:Reset noise}). For example, this is how a reset is now implemented on IBM quantum computers supported by OpenQASM3\cite{cross2021openqasm}.

In classical computing, execution of a probabilistic instruction is triggered using a random number generator (RNG), such as those widely available in software as PRNGs or in hardware as HRNGs. Here we employ a classical RNG to probabilistically activate a reset, which is identical to reset noise. In this way, artificial reset noise is implemented without ancilla qubits. Ancilla qubits would be an undesirable overhead in the larger scheme presented in this work in which unitary gates require potentially many corresponding noise channels. This hybrid approach may be viable for other noise channels. For example, reset noise can approximate amplitude damping noise to high precision \cite{gutierrez2013approximation}.

\section*{Methods}\label{sec:methodology}

\subsection*{Reservoir complexity reduction}
Reservoir complexity reduction was performed for all benchmark tasks to reduce quantum resource footprint and prevent overfitting. This involved reductions in both reservoir entanglement scheme complexities and numbers of qubits. Reduction was performed as a typical optimization procedure in which resources increase until a stopping condition is satisfied. Reductions in circuit resources was determined largely by reservoir optimization and final MSE. See Methods: Noise optimization.

Entanglement scheme complexity is quantum circuit complexity \cite{yao1993quantum}, determined by the number of entangling gates and resultant circuit depth, i.e., it is the cost of the quantum circuit. Linear entanglement schemes were trialed first for both benchmarks and were comparable to pair-separable entanglement schemes that were finally selected by the reduction principle.

The numbers of qubits in the quantum reservoirs were reduced to smaller numbers of qubits that still offered good performance. Diminishing returns were observed with reservoirs with larger numbers of qubits.

In preparation for this work, an artificial quantum noise scheme was downsized from a physical device noise model consisting of 10 submodels of thermal relaxation, depolarization, and state preparation and measurement (SPAM) noise, to a single reset noise model. A systematic reduction approach for noise channels is not presented in this work.

\subsection*{Noise optimization}
Dual annealing optimization and evolutionary optimization were employed for NARMA and Mackey-Glass benchmarks, respectively. Dual annealing from SciPy's \cite{2020SciPy-NMeth} optimization package was used for reservoir optimization using default settings. This stochastic approach, derived from \cite{xiang1997generalized}, dualizes the generalized classical simulated annealing (CSA) and fast simulated annealing (FSA) \cite{tsallis1996generalized} with a local search strategy \cite{xiang2000efficiency}. Evolutionary optimization (EO) is a population-based approach to optimization in which candidate solutions, represented as a population of agents, are initialized through random sampling. Subsequently, the fitness of each candidate solution is determined by evaluating it against a predefined objective metric. The superior solutions are then selected and utilized to generate the candidate population for the subsequent iteration. This process continues until satisfactory solutions have been identified. Here we employ the EO algorithm in Chen et al.\cite{chen2022variational}. 

Reset noise probabilities were optimized to maximize prediction performance, as detailed in the Reservoir noise parameterization. Optimization algorithms require stopping conditions. The three stopping condition were: multiple small changes in MSE, long iteration runtime without update and the maximum number of iterations was 5, which is generally observed to be a large number for the optimization algorithms used. These stopping conditions returned reproducible final MSEs, indicating that they were near optimal for the optimization algorithms.

\subsection*{Simulations}
The quantum reservoir circuits with artificial noise channels were simulated using Qiskit SDK \cite{Qiskit}. The QASM Simulator was used with an ideal density matrix simulator. This theoretical approach allows for a simulation with a single computational run of a single \texttt{QuantumCircuit} object. The single-qubit $Z$ expectation values were computed from intermediate density matrices at each time step. Simulations could not be performed with linear entanglement reservoirs larger than 12 qubits because of the demands of a density matrix simulator.

Reset noise channels are coded with \texttt{Kraus} instructions added directly to a \texttt{QuantumCircuit}. A \texttt{reset\_error} channel is available with a single probability parameter, the target of optimization. It was passed to the \texttt{Kraus} instruction.

\subsection*{Memory capacity}
Recurrency of an RC enables retention of information or a short-term memory of past signals in reservoir states. The memory capacity (MC) is a measure which quantifies this ability to retain information of the past inputs and it plays a crucial role in the prediction abilities of a reservoir computer\cite{jaeger2001echo}. 

To calculate MC, first a random sequence from a uniform distribution is prepared that is appropriate to for optimized QNIR model. The minimum and maximum values of the random sequence, i.e. the scale of the values, is made to be equivalent to benchmark time series scale that the model was optimized for. QNIR is then trained to predict signals $d$ timesteps before the input sequence of the reservoir, $u_k$, where the target signal is $\hat{y}_k = u_{k-d}$. The memory function (MF) is defined as the square of the Pearson correlation coefficient,
\begin{equation}
    M\!F_d = \frac{\text{cov}^2(y_k, \hat{y}_k)}{\sigma^2(y_k)\sigma^2(\hat{y}_k)},
\end{equation}
and the MC is then calculated as the sum of the MFs for all the delays as
\begin{equation}
    M\!C = \sum_{d}^{} M\!F_d.
\end{equation}

In Results, MCs are calculated for QNIR models that were trained and optimized for the three NARMA and two Mackey-Glass systems.

\subsection*{Metrics}
Metrics normalized mean squared error (NMSE) and normalized root-mean-square error (NRMSE) are frequently used in the relevant literature and therefore they are used here for convention and comparison. The mean absolute scaled error (MASE) metric of Hyndman and Koehler \cite{hyndman2006another} is used due to its many properties that allow for comparison between time series of different scales and is readily interpretable due to symmetry and linearity. Furthermore, MASE is used because we compare QNIR prediction performance with the Naive model, whose performance is better than a linear model and thus provides a more challenging reference prediction. The Naive model is one of the simplest forecasting models, in which the next time series value is predicted to be equal to the current value \cite{hyndman2018forecasting}.

The mean squared error (MSE) used as an optimization cost function is defined
\begin{equation}
\text{MSE} = \frac{1}{n} \sum_i (y_i - \hat{y}_i)^2.
\end{equation}
NMSE used to evaluate prediction performance is defined
\begin{equation}
\text{NMSE} = \frac{\sum_i (y_i - \hat{y}_i)^2}{\sum_i y_i^2}.
\end{equation}
NRMSE is defined
\begin{equation}
\text{NRMSE} = \frac{\sqrt{\frac{1}{T}\sum_t (y_i - \hat{y}_i)^2}}{\sigma(\mathbf{y})}
\end{equation}
where $\sigma(y)$ is the sample standard deviation of the true values.
MASE forecasting metric is defined
\begin{equation}
\text{MASE} = \frac{\frac{1}{J} \sum_j \vert{e_j}\vert}{\frac{1}{T-1} \sum_{t=2}^T \vert{y_t - y_{t-1}} \vert}
\end{equation}
where $e_j = y_j - f_j$ is the true value minus the forecasted value. The denominator is the mean absolute error (MAE) of the non-seasonal Naive out-of-sample forecast.

\section*{Results}\label{sec:results}

\subsection*{NARMA}\label{NARMA prediction}

We show that QNIR with noise optimization has excellent theoretical performance for the nonlinear auto-regressive moving average (NARMA) sequence prediction benchmarks \cite{atiya2000new, suzuki2022natural}. A NARMA regression task involves learning the nonlinear NARMA map between a fixed input sequence and a NARMA output sequence. We label the sequences NARMA$N$, where $N$ is the order of the NARMA map. We consider three NARMA sequences of orders 2, 5 and 10.

\begin{figure*}
\includegraphics[width=\linewidth]{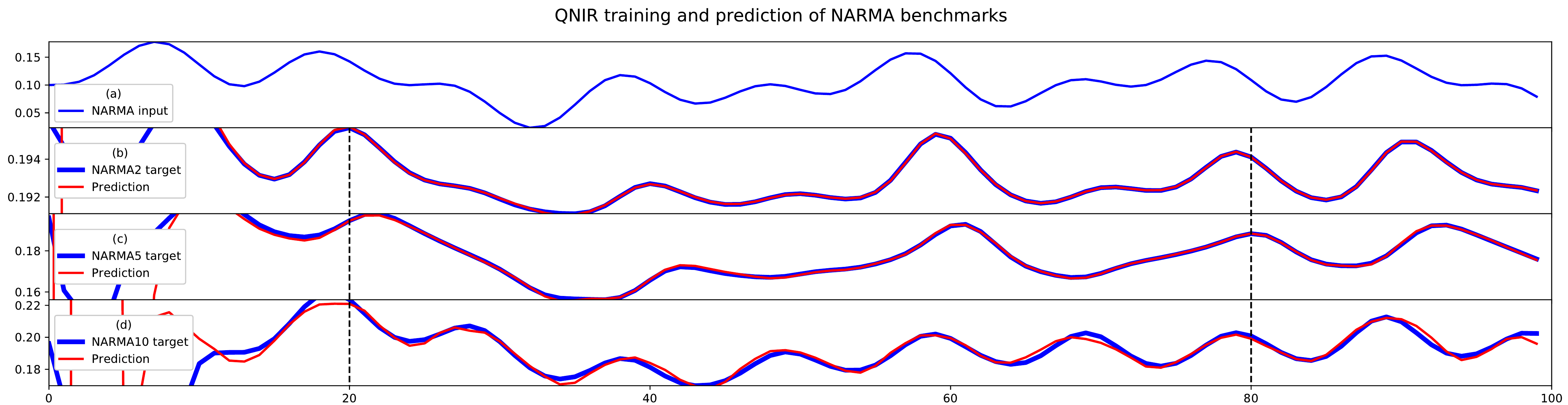}
\caption{Plot (a) is the input sequence for all NARMA tasks, Eq. \ref{eqn:input sequence u}. Plots (b-d) are QNIR training and prediction of NARMA2, 5, 10 maps, respectively. Training data sequences are between time step indices 20-80 and test data sequences are between 81-100. Test set prediction performance metrics are in Table \ref{table:1}.}
\label{fig:NARMA plots}
\end{figure*}
The NARMA2 sequence \cite{atiya2000new} is given by the recurrence relation
\begin{equation}
y_{t+1} = 0.4y_t + 0.4 y_t y_{t-1} + 0.6 u_t^3 + 0.1,
\end{equation}
where the two initial sequence values are $\{0.196, 0.19468\}$. The input values $u_t$ are from the smooth function 
\begin{equation}
u_t = 0.1 \sin \left( \frac{2\pi a t}{T} \right)\sin \left( \frac{2\pi b t}{T} \right)\sin \left( \frac{2\pi c t}{T} \right) + 0.1
\label{eqn:input sequence u}
\end{equation}
where $(a,b,c,T)=(2.11,3.73,4.11,100)$. NARMA5 and NARMA10 are described by the following general recursive function
\begin{equation}
y_{t+1} = \alpha y_t + \beta y_t \left( \sum_{i=0}^{n-1} y_{t-i} \right) + \gamma u_{t-(
n-1)} u_t + \delta.
\end{equation}

\begin{figure*}
\includegraphics[width=\linewidth]{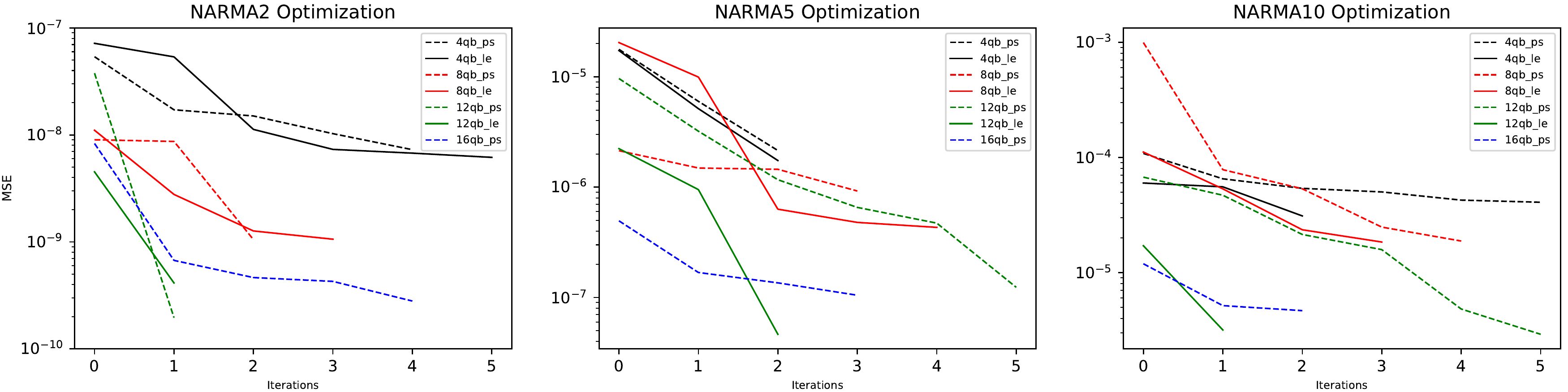}
\caption{Log plots of MSE cost curves for reservoir tuning iterations for both pair-separable (PS) and linear entanglement (LE) reservoir designs. The number of qubits in both PS and LE reservoirs were increased from 4 qubits in steps of 4 qubits until 12-qubit reservoirs were selected. For NARMA2 and 10 both PS and LE reservoirs had similar final MSE values. For NARMA5 the LE reservoir provided a better option. See Methods for details on optimization and simulation matters.}
\label{fig:NARMA_optimization_plots}
\end{figure*}

For NARMA5, the initial sequence is $\{0,0,0,0,0.196\}$ and the first four zeroes are excluded from the target sequence. The function parameters for NARMA5 are $(\alpha,\beta,\gamma,\delta,T)=(0.3, 0.05, 1.5, 0.1, 100)$. Similarly for NARMA10, the first nine values in the initial sequence are zeroes and are not included in the target sequence. The function parameters used are the same as for NARMA5. NARMA time series values were encoded directly to the angle of the encoding gates. Temporal train and test split indices are 20-80 and 81-100, respectively. The initial 20 time steps were excluded as a washout phase.
\begin{table}[h!]
\small
\begin{center}
\begin{tabular}{ |p{2cm}||p{2cm}|p{2cm}|p{2cm}|  }
 \hline
 \multicolumn{4}{|c|}{QNIR NARMA Results} \\
 \hline
 Metric & NARMA2 & NARMA5 & NARMA10\\
 \hline
 MASE   & $2.8 \times10^{-2}$ & $6.4 \times10^{-2}$  &  $0.39$ \\
 NMSE & $6.1 \times10^{-9}$ & $1.4\times10^{-6}$ & $8.3 \times10^{-5}$\\
 NRMSE & $1.6 \times10^{-2}$ & $3.3 \times10^{-2}$ & $0.22$\\
 MSE & $2.3 \times10^{-10}$ & $4.6 \times10^{-8}$ & $3.2 \times10^{-6}$\\
 \hline
 \multicolumn{4}{|c|}{Naive model NARMA Results} \\
 \hline
  MASE   & $1$ & $1$ & $1$\\
 NMSE & $6.5\times 10^{-6}$ & $3.0\times 10^{-4}$ & $6.1\times 10^{-4}$ \\
 NRMSE & $0.51$ & $0.48$ & $0.60$ \\
 \hline
\end{tabular}
\end{center}
\caption{QNIR performance metrics are explicitly compared with the Naive model, which has an out-of-sample MASE of 1 by definition (Methods: Metrics). NARMA2 and 10 were achieved with pair-separable reservoir and NARMA5 with linear entanglement.}
\label{table:1}
\end{table}
\\
\\
Excellent simulation results have been achieved for the NARMA2, 5 and 10 tasks, plotted in Fig. \ref{fig:NARMA plots} and recorded in Table \ref{table:1}. The primary reason for the high quality results comes from the effectiveness of the reset noise parameterization and subsequent optimization, first implemented in this work. This effectiveness is demonstrated by a 2 orders of magnitude improvement from random initialization of reset noise probabilities to the final optimization iteration for all three NARMA tasks, as plotted in Fig. \ref{fig:NARMA_optimization_plots}. For each of the three NARMA tasks, three distinct sets of optimal parameters were obtained. Information processing capacity (IPC) analysis \cite{dambre2012information} has been used to show that QNIR can solve the NARMA2 task \cite{kubota2022quantum}. The excellent performance on NARMA2 in particular provides evidence that noise parameterization is suitable for approaching an optimal solution model.

In combination with reservoir noise optimization, reduction of the quantum reservoirs was performed in number of qubits and reservoir circuit complexity in terms of entanglement schemes (see Methods). 
Pair-separable (PS) and linear entanglement (LE) scheme-based reservoirs were optimally reduced to 12 (6$\times$2) qubits for all NARMA tasks. Reservoirs with larger number of qubits did not improve NARMA prediction performance and smaller numbers of qubits show a drop in performance, see Fig. \ref{fig:NARMA_optimization_plots}. 12-qubit reservoirs are parameterized with 42 and 67 reset noise probabilities for PS and LE reservoirs, respectively. Next we consider the second dimension of reduction of entanglement schemes. For NARMA 2 and 10, the reservoir complexity and associated entanglement was reduced to a PS reservoir. This is because the results for LE reservoirs were comparable. However, for NARMA5 an LE reservoir was not reduced because of improved performance. The quantum state of the LE reservoir is non-separable due to a higher degree of entanglement.

\begin{figure*}
\includegraphics[width=\linewidth]{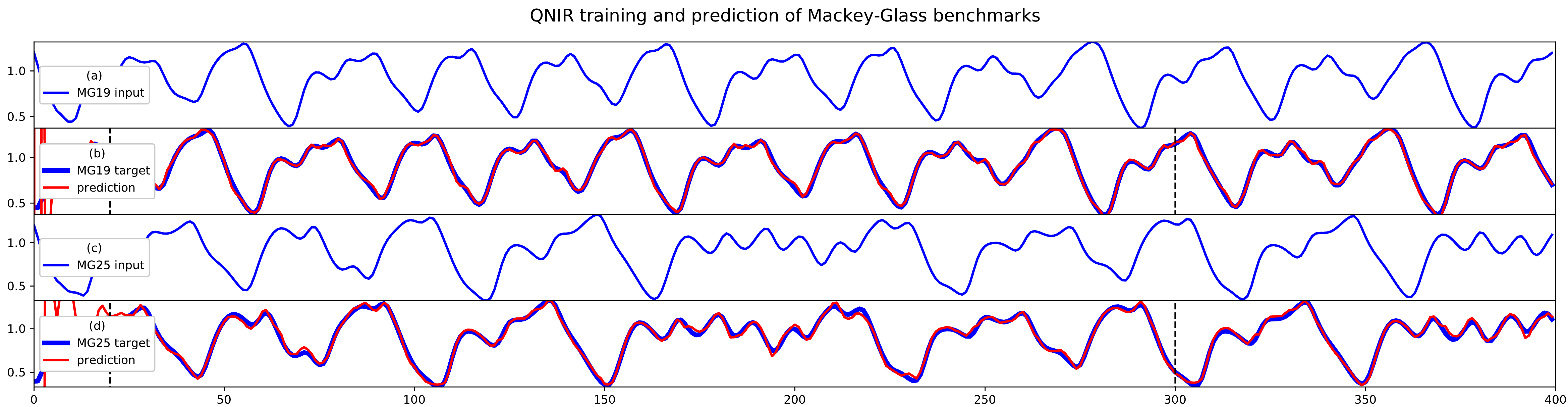}
\caption{The input and target time series are defined as $x(t-\tau)$ and $x(t)$, respectively (Eq. \ref{eqn:Mackey-Glass equation}). Plots (a-b) are MG19 input and 19-step delay target time series and prediction result, respectively. The same applies for plots (c-d) for MG25. Training data sequences are between time step indices 20-300 and test data sequences are between 301-400. Test set prediction performance metrics are in Table \ref{table:2}.}
\label{fig:MG plots}
\end{figure*}

This result strongly indicates an improvement from recent work \cite{suzuki2022natural} in terms of reducing resource-noise requirements to a single reset noise model. Those experimental results \cite{suzuki2022natural} have inherent measurement sampling error, however, our result demonstrates that multifaceted physical device noise is not required for the NARMA tasks defined in these works, as only reset noise channels were required here. These results may further suggest that a reset noise QNIR would be a favorable direction for more NARMA benchmarks \cite{kubota2022quantum}. 

MCs of the systems were observed to be saturated at $4.55(\pm0.07)$, $4.50(\pm0.07)$, and $4.70(\pm0.06)$ at the delays of 8, 8, and 10 for NARMA2, 5, and 10, respectively (see Fig. \ref{fig:NARMA-MG_MF_plots}). Confidence intervals at $\alpha=0.05$ are indicated in brackets. The MF profiles differ from unoptimized QNIR reservoirs, which suggests structural differences in short-term memory. Unoptimized reservoirs have more sigmoid or S-shaped curves with lower MC. The NARMA10 MF plot has very clear structure in MF values that are larger for longer delays up to 10, needed for the higher order function. A partial reason for the lower NARMA10 prediction result may be the smaller MF value at $d=10$.
The memory-nonlinearity trade-off inherent in RC algorithms \cite{inubushi2017reservoir} should be established and investigated in a dedicated work for QNIR to aid interpretation of these metrics.

\subsection*{Mackey-Glass}

The Mackey-Glass (MG) nonlinear system \cite{mackey1977oscillation} is a commonly used benchmark for time series prediction that is difficult to predict due to challenging chaotic dynamics under some parameters. The Python package ReservoirPy \cite{trouvain2020reservoirpy} was used to generate MG system time series, which are discretized using the Runge-Kutta method and initialized with a default seed value. For MG benchmarking, we extend the training sequence from 60 to 250 data points and the testing sequence from 20 to 100 data points from what was used for NARMA. This extension is designed to stress test QNIR.

The MG delay differential equation (DDE) is
\begin{equation}
\frac{dx}{dt} = \frac{ax(t-\tau)}{1+x(t-\tau)^n} - bx(t).
\label{eqn:Mackey-Glass equation}
\end{equation}
To generate time series for benchmarking, parameters $(x_0,a,b,n) = (1.2,0.2,0.1,10)$ were used. The input and target time series are defined as $x(t-\tau)$ and $x(t)$, respectively. We considered two distinct, chaotic MG systems determined by integer delay values $\tau=19$ and $25$, which we denote MG19 and MG25, respectively. The generated time series were then downsampled by a factor of 2. For both downsampled MG19 and MG25 time series, chaoticity is indicated by positive Lyapunov exponents \cite{eckmann1986liapunov}, calculated using the \emph{nolds} Python library \cite{scholzel2019nonlinear}.

The time series were downsampled from $800$ time steps to $400$ time steps. Temporal train and test split indices are 20-300 and 301-400, respectively. The initial 20 time steps were excluded as a washout phase. The MG time series values were encoded directly to the angle of the encoding gates.

\begin{table}[h!]
\small
\label{table:QNIR NARMA Results}
\begin{center}
\begin{tabular}{ |p{2cm}||p{2.5cm}|p{2.5cm}|  }
 \hline
 \multicolumn{3}{|c|}{QNIR Mackey-Glass Results} \\
 \hline
 Metric & MG19 & MG25\\
 \hline
 MASE   & $0.29$ & $0.38$\\
 NMSE & $4.6 \times10^{-4}$ & $7.3 \times10^{-4}$\\
 NRMSE & $8.7 \times10^{-2}$ & $0.10$\\
 MSE & $4.3 \times10^{-4}$ & $6.5 \times10^{-4}$\\
 \hline
 \multicolumn{3}{|c|}{Naive model Mackey-Glass Results} \\
 \hline
  MASE   & $1$ & $1$ \\
 NMSE & $4.4 \times10^{-3}$ & $4.0 \times10^{-3}$\\
 NRMSE & $0.27$ & $0.24$\\
 \hline
\end{tabular}
\end{center}
\caption{QNIR performance metrics are explicitly compared with the Naive model, which has an out-of-sample MASE of 1 by definition (Methods: Metrics).}
\label{table:2}
\end{table}

\begin{figure*}
\includegraphics[width=\linewidth]{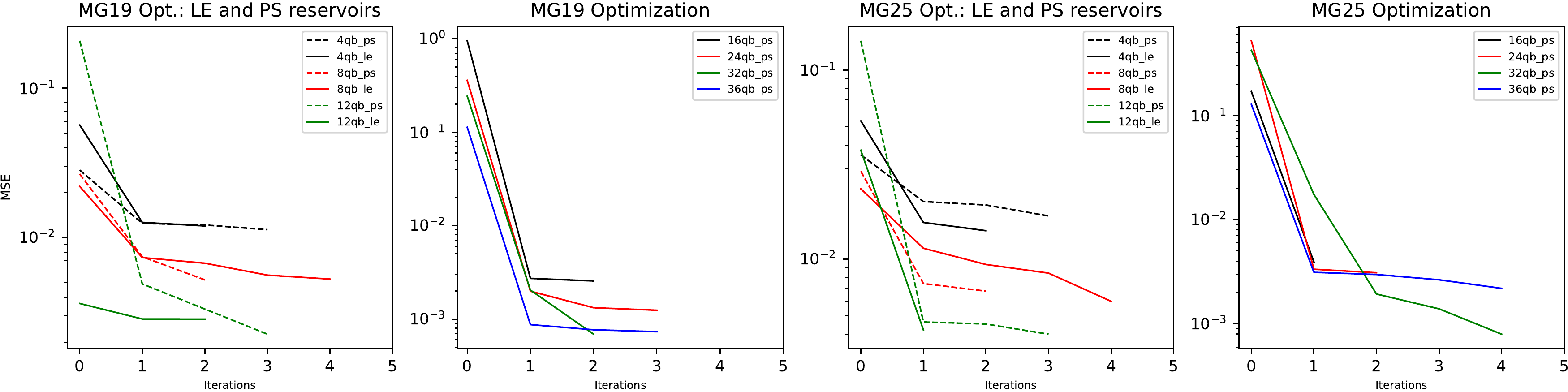}
\caption{Log plots of MSE cost curves for reservoir tuning iterations. MSEs from initial to final iterations improved up to 2.5 orders of magnitude, demonstrating the effectiveness of noise optimization. In the two plots comparing LE and PS reservoirs, it can be seen that there is no notable difference between final MSEs for MG19 or MG25 data up to 12 qubits. By the principle of resource reduction PS reservoirs should be favored. See Methods for details of optimization and simulation matters.}
\label{fig:MG optimization plots}
\end{figure*}

\begin{figure}[h!]
\centering
\includegraphics[width=13cm]{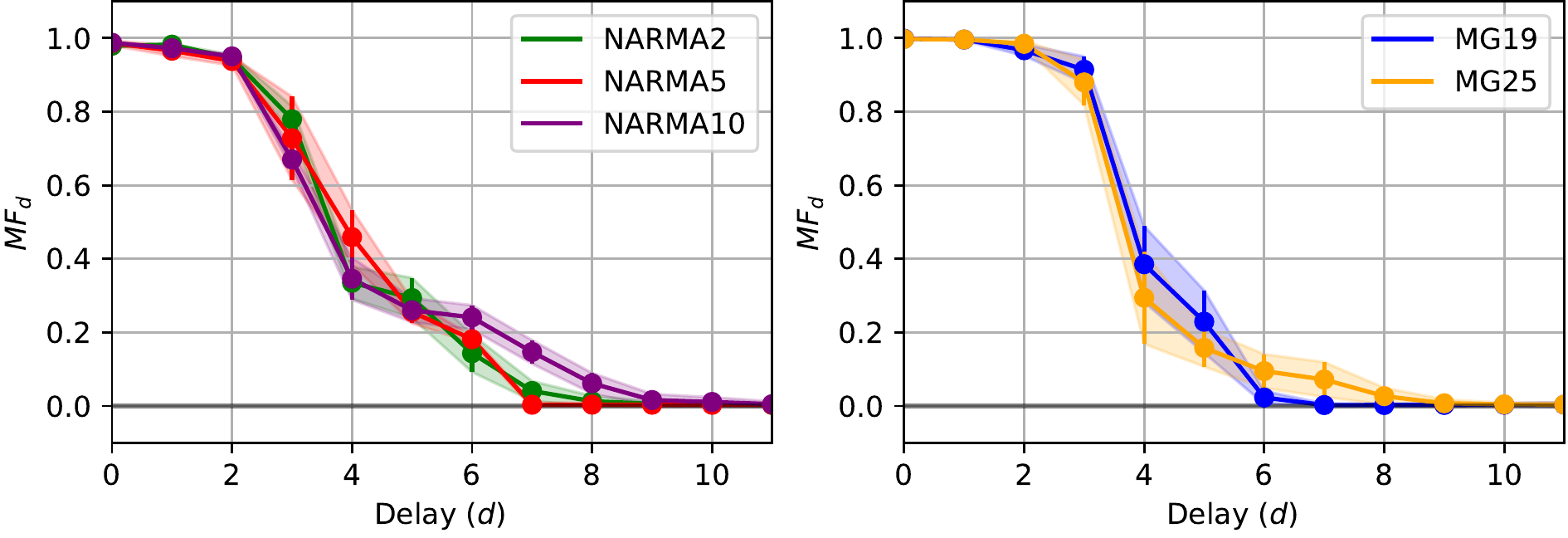}
\caption{QNIR memory functions for NARMA and Mackey-Glass tasks over 30 trials plotted against delay $d$ in the input signal. The colored bands correspond to the standard deviations.}
\label{fig:NARMA-MG_MF_plots}
\end{figure}

We report good prediction performances, plotted in Fig. \ref{fig:MG plots} and recorded in Table \ref{table:2}. QNIR has demonstrated prediction performances much better than the Naive model and shows promise for modeling challenging chaotic dynamics with exponential sensitivity to initial conditions. Larger reservoirs with three times the number of qubits were required for the MG modeling compared to NARMA, indicating greater prediction difficulty. However, it is worth emphasizing that 32-qubit reservoirs are still relatively small by conventional approaches.

The effectiveness of noise parameterization and optimization can be seen in Fig. \ref{fig:MG optimization plots} with the large initial drops from iteration 0 to 1, where iteration 0 reflects randomly initialized parameters. This is a main result in demonstrating the effectiveness of this noise parameterization approach.

By the reduction procedure, reservoirs were reduced to 32 (16$\times$2) qubits for MG19 and MG25 tasks. Increasing the number of qubits beyond these numbers saw diminishing returns and reductions below caused a drop-off in performance, as can be seen in Fig. \ref{fig:MG optimization plots}. The 32-qubit reservoirs were parameterized with 112 noise probabilities. Comparable performances were obtained for both LE and PS reservoirs, therefore by reduction PS reservoirs were selected.

MCs of the systems were saturated at $4.56(\pm0.07)$ and $4.55(\pm0.08)$ at the delays of 6, and 8 for MG19 and MG25, respectively (see Fig.~\ref{fig:NARMA-MG_MF_plots}). Confidence intervals at $\alpha=0.05$ are indicated in brackets. Since MCs for these larger MG reservoirs were similar to those utilized for the NARMA benchmark, the larger modeling complexity may be provided by the threefold number of available reservoir output signals to the linear regression layer. The MF plots suggest a small memory design for these QNIR reservoirs although an in-depth analysis of MG system would be required to confirm. Further investigation would center on memory-nonlinearity trade-off, which may explain why smaller memories were traded for greater reservoir nonlinearity for the chaotic MG time series.

\section*{Conclusions}\label{sec:conclusion}
We have demonstrated a new QNIR reservoir optimization approach that uses parameterized resource noise to address the need for quantum reservoir tuning for improved prediction performance. This parameterization approach to reservoir tuning embodies a new, general quantum circuit parameterization approach for QML models.

Benchmarking has demonstrated that resource noise parameterization, and optimization with dual annealing and evolutionary algorithm, is effective for improving prediction performance. Our simulations showed that few-qubit QNIR computers are capable of predicting nonlinear dynamics including challenging chaotic dynamics in the Mackey-Glass system. We demonstrated a significant minimization of noise resource over previous QNIR work, resulting in a single reset noise model being selected for the benchmark samples chosen in this work.

Systematic reduction of quantum resources in the number of qubits and entanglement scheme of the reservoir was employed. While reduction of entanglement scheme complexity may produce quantum circuits that are efficient to compute classically, this process is desirable when the learning task does not require quantum advantage. This is consistent with the machine learning principles of model selection and resource reduction. Furthermore, the QNIR framework is consistent with complex entanglement schemes, and therefore opens a path towards investigating quantum advantage.

Reducing quantum circuit complexity has positive implications for quantum hardware efficiency, which is critical for current quantum computers hindered by noise. Therefore, we recommend implementation on current quantum computers using error mitigation.

\bibliography{sample}

\section*{Acknowledgements}
We would like to thank the following people for insightful and constructive discussions, suggestions, feedback and support: Dimitris Alevras, Constantin Gonciulea, Luke Govia, Alma Ionescu, Vaibhaw Kumar, Muyuan Li, Antonio Mezzacapo, Ryutaro Ohira, Jae-eun Park, Laura Schleeper, Charlee Stefanski, Francesco Tacchino, Anna Topol, Rukhsan Ul-Haq, Kavitha Yogaraj, and the IBM Quantum team. We would like to thank Yudai Suzuki for an initial technical discussion.

\section*{Author contributions statement}
Conceptualization, D.F., A.D., S.Y.-C.C., V.R., V.M.; methodology, D.F., A.D., S.Y.-C.C., V.R., V.M.; software, D.F., S.Y.-C.C., V.R.; validation, D.F., S.Y.-C.C., V.R., V.M.; writing–original draft preparation, D.F., A.D., S.Y.-C.C., V.R.; writing–review and editing, D.F., A.D., S.Y.-C.C., V.R., V.M.; visualizations, D.F., A.D., V.R.

\section*{Ethics declaration}
\subsection*{Competing interests}
The authors declare no competing interests.

\section*{Data availability}
The datasets used and/or analyzed during the current study are available from the corresponding author on reasonable request.

\section*{Additional information}
The views expressed in this article are those of the authors and do not represent the views of Wells Fargo. This article is for informational purposes only. Nothing contained in this article should be construed as investment advice. Wells Fargo makes no express or implied warranties and expressly disclaims all legal, tax, and accounting implications related to this article.

\end{document}